\documentclass[10pt,english]{article}
\usepackage{subfigure}
\usepackage{graphicx}
\usepackage{float}
\usepackage[T1]{fontenc}
\usepackage[latin9]{inputenc}
\usepackage[margin=1.25in]{geometry}
\usepackage{float}
\usepackage{amsmath}
\usepackage{amsbsy}
\usepackage{setspace}
\usepackage{amssymb}
\usepackage{esint}
\usepackage{cite}
\newfloat{algorithm}{tbp}{loa}
\floatname{algorithm}{Algorithm}
\usepackage{hyperref}
\usepackage{listings}

\newlength\myindent
\makeatletter

\floatstyle{ruled}
\newfloat{algorithm}{tbp}{loa}
\floatname{algorithm}{Algorithm}

\usepackage{babel}

\begin{document}

\title{Persuasion, Betrayal and Regret in Election Campaigns}

\author{M. Andrecut}

\date{June 21, 2024}

\maketitle
{

\centering Unlimited Analytics Inc.

\centering Calgary, Alberta, Canada

\centering mircea.andrecut@gmail.com

} 

\smallskip 

\begin{abstract}

Elections play a fundamental role in democratic societies, however they are often characterized by unexpected results. 
Here we discuss an election campaign model inspired by the compartmental epidemiology, and  
we show that the model captures the main characteristics of an election campaign: persuasion, betrayal and regret. 
All of these three factors can be used together or independently to influence the campaign, and to determine the winner. 
We include results for both the deterministic and the stochastic versions of the model, 
and we show that the decision to not vote significantly increases the fluctuations in the model, amplifying the chance of controversial results.  

\smallskip 



\end{abstract}

\smallskip 

\section{Introduction}

Elections are some of the most important decision events in democratic societies, however they are often characterized by unexpected and contradictory outcomes \cite{key-1}, \cite{key-2}. 
It is frequently assumed that such contradictory, and sometimes counter-intuitive results are a consequence of the susceptibility of the voting procedures to various paradoxes \cite{key-3}. 

An interesting example is provided by the "paradox of not voting", which states that in a large election, the probability that an individual vote might change the election outcome is vanishingly small \cite{key-4}. 
In this particular case, the rational-choice model, which is frequently used in economics, can be used to provide some insights \cite{key-5}. 
According to the rational-choice model, in a two candidates election a voter must decide to vote for one of the candidates or to abstain. 
Assuming that $B>0$ is the benefit to the voter that the chosen candidate wins, $C>0$ is the cost of voting, and $p\in [0,1]$ is the probability that 
this single vote will change the outcome of the election in favor of the chosen candidate, then the voter should vote rather than abstain if and only if $p>C/B$. 
Since the probability that a single vote might change the election outcome is vanishingly small, we anticipate that very few people would have an incentive to vote. 

Most people also assume that abstention from vote affects the candidates in equal measure. However, this assumption is also false since 
in fact, abstention could work in favor (or against) of one of the candidates \cite{key-6}. 
For example, if the voters help their chosen candidate by withholding their vote, then this phenomenon is known as the "no-show paradox" \cite{key-7}.  

We can see that the sole economic or rational reasons are not enough to understand why people vote or abstain in elections. 
Obviously, the reasons to vote should exceed simple economics, otherwise as shown by the "paradox of not voting", nobody would have any real incentives to vote. 
However, in reality many people are still making the effort to cast their vote, suggesting that the voters are not rational. 

To fully understand why and how people vote, it is very likely necessary to go beyond the simplistic economics ideas, and to also approach the problem from a different angle, such as the social psychology \cite{key-8}. 
For example, a more recent study \cite{key-9}, identifies five motives for voting turnout: selfishness, altruism, personal duty, belonging, and social approval. 
This study also shows that duty and selfishness are the most important reported motives for voting turnout. While such studies are more or less successful in explaining the motives behind the voting turnout, 
they do not take into account the time component, and the dynamics of the election campaigns.
Therefore, they do not explain how the voters are actually influenced during the finite time of the election campaigns, by the actions (strategy) of the candidates, and their hard core supporters. 

Motivated by the above observations, here we discuss an election campaign model inspired by the compartmental epidemiology \cite{key-10}, \cite{key-11},  
and we show that the proposed model captures some of the main characteristics of an election campaign: persuasion, betrayal and regret. 
We show how these three characteristics of the model can be used together or independently to influence the campaign outcome, and to determine the winner during the finite time of the campaign. 
We include numerical results for both the deterministic and the stochastic versions of the model, and we show that the decision to not vote (abstention) increases the fluctuations in the model, 
amplifying the chance of controversial results. 

\section{Deterministic model}

\subsection{Epidemic inspired election model}
As previously mentioned in the introduction, our election campaign model is inspired by the compartmental epidemic models \cite{key-11}. 
The compartmental epidemic models are used to analyze the disease dynamics, and assume that the population can be categorized into several distinct types, or "compartments". 
The contagion and recovery dynamics is described by the transitions of the population individuals between these compartments. 

In particular, our election campaign model is inspired by the susceptible-infected-susceptible (SIS) epidemic model, where a population of $N$ individuals is divided in two compartments, such that 
at a given time $t$ an individual can be either susceptible (S) or infected (I) \cite{key-11}. 
The susceptible and infected populations depend on two parameters: the infection transmission rate $\beta$, and the recovery rate $\delta$. 
This means that a susceptible individual becomes infected with the rate $\beta$ when it interacts with an infected individual, and respectively 
an infected individual is recovering with the rate $\delta$, becoming susceptible again. 
Therefore, following the above assumptions, the SIS epidemic model can be described using the following set of 
ordinary differential equations:
\begin{align}
\frac{dS}{dt} &= -\frac{\beta}{N} SI + \delta I,\\
\frac{dI}{dt} &= \frac{\beta}{N} SI - \delta I,\\
N &= S + I.
\end{align}
Here, the third equation is the total population conservation constraint.

Epidemic models, such as the SIS model, are typically formulated to describe the dynamics of a single disease. 
Our election campaign model extends this approach to the case where two (or more) diseases are competitively spreading through the susceptible population. 

Let us assume a population of $N$ individuals, where two of them are candidates for an election. 
We denote by $x$ the population supporting the first candidate, and by $y$ the population supporting the second candidate. 
Also, we denote by $s$ the susceptible population. 
Initially at $t=0$ we assume that $x_0 \geq 1$, and $y_0 \geq 1$, such that $(x_0,y_0)$ can be identified as the initial "core" supporters of the two candidates. The election 
campaign consists of a fixed allocated time interval $[0,T]$, where the $x$ and $y$ supporters are trying to persuade the susceptible population $s$ to 
vote for their candidate. In order to simplify the model description, and to maintain clarity, here we limit our investigation to only two candidates. 
Using the above epidemic analogy, the deterministic election model can be written as the following system of differential equations:
\begin{align}
\frac{ds}{dt} &= -\frac{\beta_x}{N} sx -\frac{\beta_y}{N} sy + \delta_x x + \delta_y y,\\
\frac{dx}{dt} &= \frac{\beta_x}{N} sx - \delta_x x + \frac{\eta}{N}xy,\\
\frac{dy}{dt} &= \frac{\beta_y}{N} sy - \delta_y y - \frac{\eta}{N}xy,\\
N &= s + x + y.
\end{align}
Here, $\beta_{x,y}\geq0$ are the persuasion rates, $\delta_{x,y}\geq 0$ are the regret rates, and $\eta$ is the net betrayal rate. 
That is, $x$ and $y$ are successfully convincing $s$ to join them with the rates $\beta_{x,y}$, 
and respectively $x$ and $y$ regret their choice and change their mind, becoming susceptible again with the rates $\delta_{x,y}$. 
One can see that these two processes are analogous to the infection (persuasion) and recovery (regret) processes in the SIS epidemic model. 
In addition, we also assume that the members of $x$ and $y$ can betray (change sides) with a net rate $\eta$ under the influence of the other side. 
The net betrayal rate $\eta$ can be positive or negative: $\eta>0$ if $y$ betrays more than $x$, and $\eta<0$ vice versa. 
The fourth equation is the population conservation constraint, which normalizes the results to the total number of voters. 
We should emphasize here that in this model, the regret can be also interpreted as the not voting decision. 

\subsection{No betrayal, and no regret}

Let us first consider the case when both $x$ and $y$ have no regret in choosing their candidate and they also do not betray, which means that  $\delta_{x,y} = 0$ and $\eta=0$.  
In this case the model becomes:
\begin{align}
N \frac{ds}{dt} &= -\beta_x sx -\beta_y sy,\\
N \frac{dx}{dt} &= \beta_x sx,\\
N \frac{dy}{dt} &= \beta_y sy,\\
N &= s + x + y.
\end{align}

One can easily see that by eliminating $s$ from the second and third equations we obtain:
\begin{equation}
\beta_y \frac{dx}{x} = \beta_x \frac{dy}{y},
\end{equation}
which can be integrated as following:
\begin{equation}
\beta_y \int^x_{x_0} \frac{dx}{x} = \beta_x \int^y_{y_0} \frac{dy}{y},
\end{equation}
\begin{equation}
\beta_y (\ln x - \ln x_0) = \beta_x (\ln y - \ln y_0).
\end{equation}
From here we obtain:
\begin{equation}
y = y_0 \left( \frac{x}{x_0} \right)^\xi,
\end{equation}
where $\xi=\beta_y/\beta_x$. 
Since we always have $x,y>0$, the steady state $(x_*$, $y_*)$ is reached when $s=0$, which means:
\begin{equation}
x_* + y_* = N.
\end{equation}
By inserting (15) into (16) we obtain a transcendental equation for $x_*$:
\begin{equation}
x_* + y_0 \left( \frac{x_*}{x_0} \right)^\xi = N,
\end{equation}
which can be solved numerically to find the steady state.

The gap of the steady state solution is a nonlinear function:
\begin{equation}
\Delta(x_*,\alpha,\xi) = x_* - y_* = x_* -  y_0 \left( \frac{x_*}{x_0} \right)^\xi.
\end{equation}
Thus, the winner will be dictated by the initial condition (the "core" support) $(x_0,y_0)$ and the parameter $\xi$, which is the ratio of the persuasion rates. 

Let us assume that $x$ wins with a gap $\Delta > 0$, then we have:
\begin{equation}
x_* = \frac{N+\Delta}{2}, \quad y_* = \frac{N-\Delta}{2},
\end{equation}
and from the gap equation we obtain the required $\xi$ value in order to achieve the desired gap $\Delta$:
\begin{equation}
\xi =\frac{\beta_y}{\beta_x} = \frac{\ln \left( \frac{N-\Delta}{2y_0} \right)}{\ln \left( \frac{N+\Delta}{2x_0} \right)}
\end{equation}

In Figure 1 we illustrate these results by considering a population of $N=10^6$, with equal initial conditions $x_0=y_0=100$ ("core support"). We also assume that the persuasion parameter of $y$ is $\beta_y = 0.25$. 
If $x$ would like to win with a gap of $\Delta = 10^5$ ($10\%$ of the total population), then it needs a persuasion coefficient of $\beta_x=0.255963941$ (Figure 1(a)). 
However, if the initial core support for $x$ drops to $x_0=75$ then, with the same $\beta_x$, $x$ will lose with a gap of $\Delta \simeq -41736$ (Figure 1(b)). 

This particular case emphasizes how important is the initial "core" support of each candidate in an election, since a larger initial core allows for a faster growth of the number of supporters during the campaign, 
in the absence of betrayal and regret.

\begin{figure}[!t]
\centering \includegraphics[width=7.56cm]{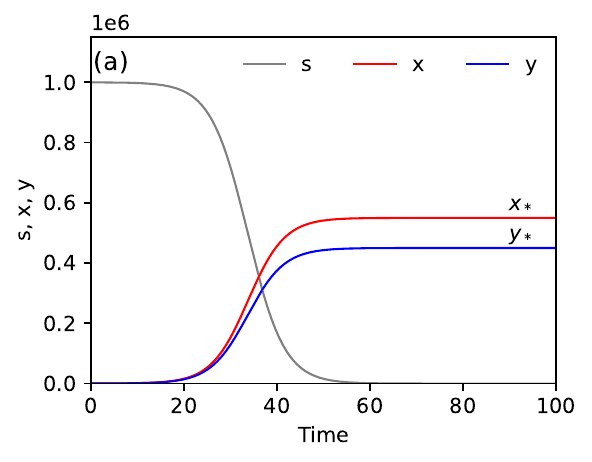}
\centering \includegraphics[width=7.56cm]{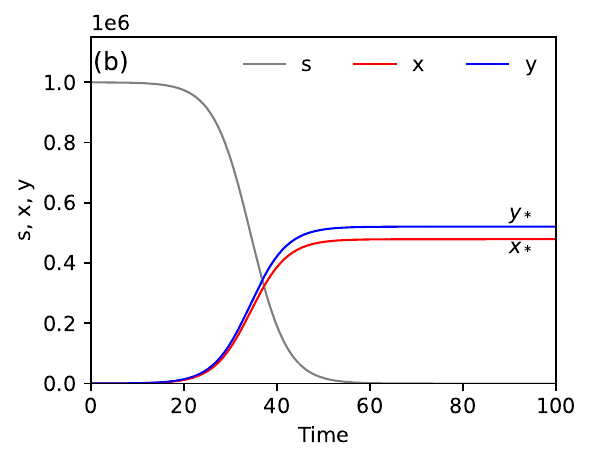}
\caption{Simulation results for the deterministic model with no betrayal and no regret. The parameters are $N=10^6$, $\beta_x=0.255963941$, $\beta_y = 0.25$, $\delta_x=0$, $\delta_y=0$, $\eta=0$ and: (a) $x_0=y_0=100$ (b) $x_0=75, y_0=100$.}
\end{figure}

\subsection{Regret, but no betrayal}

Let us now consider the case when we have no betrayal, and therefore $\eta=0$, but regret is present in the system $\delta_{x,y} > 0$,  
such that the system becomes:
\begin{align}
\frac{ds}{dt} &= -\frac{\beta_x}{N} sx -\frac{\beta_y}{N} sy + \delta_x x + \delta_y y,\\
\frac{dx}{dt} &= \frac{\beta_x}{N} sx - \delta_x x,\\
\frac{dy}{dt} &= \frac{\beta_y}{N} sy - \delta_y y,\\
N &= s + x + y.
\end{align}
This model is basically equivalent to an SIS epidemic model, where two diseases are competing on the same susceptible population, 
but a susceptible individual can be infected by only one of the diseases at a given time. Once the individual recovers it becomes 
susceptible again, and eventually can be re-infected by the same disease, or by the other disease, if the limited time of the election campaign allows it. 

In this case, the steady states of the differential equations system can be obtained by solving the following non-linear system of equations:
\begin{align}
\frac{\beta_x}{N} sx - \delta_x x &= 0,\\
\frac{\beta_y}{N} sy - \delta_y y &= 0,\\
s + x + y &= N.
\end{align}

\begin{figure}[!t]
\centering \includegraphics[width=7.56cm]{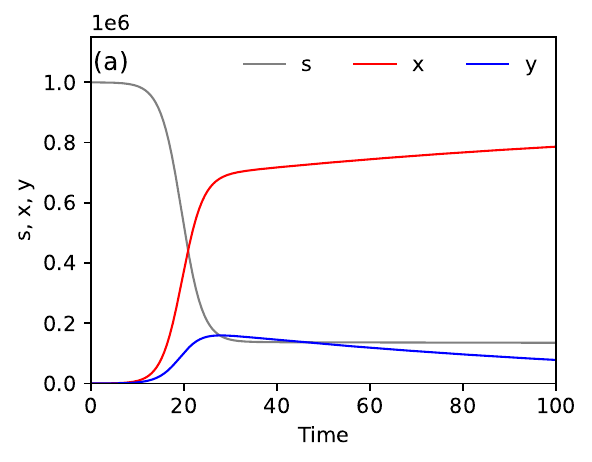}
\centering \includegraphics[width=7.56cm]{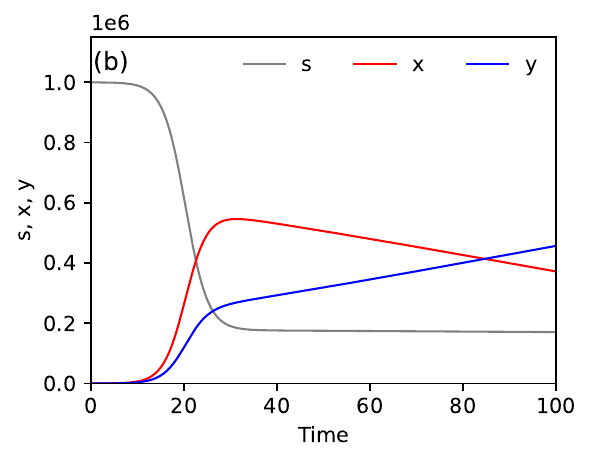}
\caption{Simulation results for the deterministic no betrayal model with increasing regret. The parameters are $T=100$, $N=10^6$, $x_0=100$, $y_0=75$, $\beta_x=0.525$, $\beta_y=0.475$, $\eta=0$, $\delta_y=0.075$ and: 
(a) $\delta_x=0.07$; (b) $\delta_x=0.097$.}
\end{figure}

One can see that we always must have $s_*\geq 0$, and therefore either or both $x_*$ and $y_*$ must be equal to zero. 
Thus, besides the trivial steady state $(x_*,y_*,s_*)=(0,0,N)$, we can have other two possible steady states, depending on the values of the model parameters:
\begin{align}
x_*^{(1)} = N\left( 1- \frac{\delta_x}{\beta_x} \right), \quad y_*^{(1)} &= 0, \quad s_*^{(1)} = N\frac{\delta_x}{\beta_x},
\end{align}
and respectively
\begin{align}
x_*^{(2)} = 0, \quad y_*^{(2)} = N\left( 1- \frac{\delta_y}{\beta_y} \right), \quad s_*^{(2)} = N\frac{\delta_y}{\beta_y}.
\end{align}
It is interesting to note that none of these steady states can be actually reached, unless the candidates vote against themselves. 

Also, we should note that the steady states of the system do not depend on the values of the initial condition $x_0 \geq 1$, $y_0 \geq 1$ (the initial supporting "core"). 
However, the speed of approaching the steady states does depend on the initial condition values, and therefore it is not immediately obvious that a given 
set of $\beta$ and $\delta$ parameters will guarantee the winning of the election in the fixed time length $T$ allocated to the campaign. 
Thus, as mentioned in the introduction, the campaign time length also plays an important role if regret (not voting) is included in the model, because the candidate that will win in a shorter campaign may unexpectedly lose at the end of a longer campaign, 
due to a faster increasing regret of the voters. Unfortunately, a closed form solution of the differential equations system is not known, and therefore numerical simulations are necessary 
to illustrate these observations.

In Figure 2 we show such a numerical result, where the length of the campaign is limited to $T=100$. By increasing the regret (not voting) rate $\delta_x$ and keeping the rest of the parameters constant,  
the election result is changing dramatically for the candidate $x$, from a categoric win in Figure 2(a), to a loss in Figure 2(b), even though $x$ had an advantage for most of the campaign, 
due to a greater persuasion rate and a better "core" support in the initial condition. 
However, because in the second case the population increasingly regrets the initial voting "impulse", it can unexpectedly lose the election by the end of the campaign.

Figure 2 also shows that a higher regret rate induces a significant increase in the not voting (susceptible) population. 
In Figure 3 we exemplify this observation by doubling the regret rates. In Figure 3(a) the regret rates are $\delta_x=0.09$, $\delta_y=0.075$,  
and the $x$ candidate wins with a comfortable large gap in the campaign time $T=100$. However, in Figure 3(b) the values of the regret rates are doubled, $\delta_x=0.18$, $\delta_y=0.15$, while the rest of the parameters are kept the same, 
resulting in doubling the number of the voting absentees, and consequently in an unexpected loss for the candidate $x$.

\begin{figure}[!t]
\centering \includegraphics[width=7.56cm]{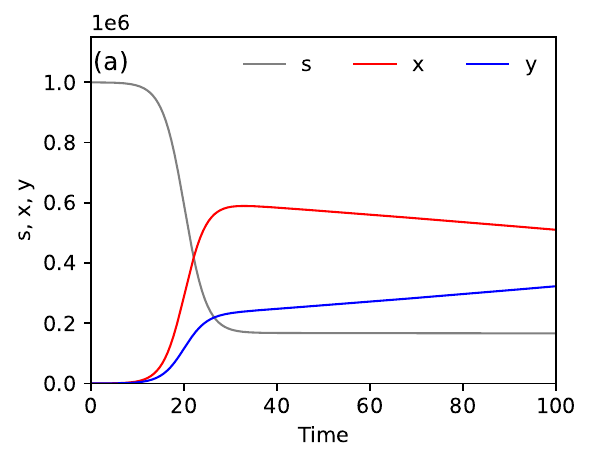}
\centering \includegraphics[width=7.56cm]{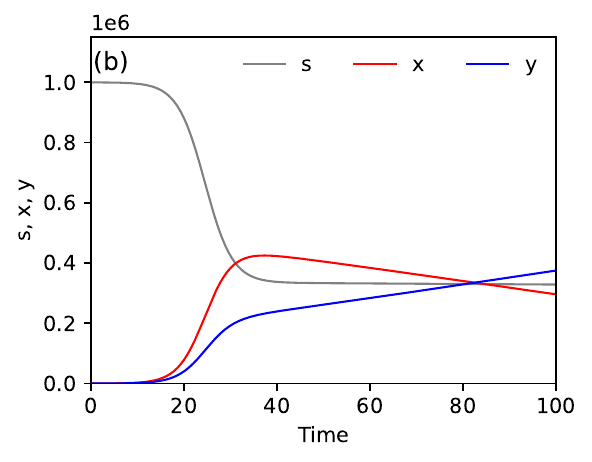}
\caption{Not voting effect in the deterministic model with regret, but no betrayal. The parameters are $T=100$, $N=10^6$, $x_0=100$, $y_0=75$, $\beta_x=0.525$, $\beta_y=0.475$, $\eta=0$: 
(a) $\delta_x=0.09$, $\delta_y=0.075$; (a) $\delta_x=0.18$, $\delta_y=0.15$.}
\end{figure}

\subsection{Betrayal, but no regret}

Let us now assume the case with a nonzero betrayal rate $\eta\neq0$, and no regret $\delta_x=\delta_y=0$, such that we have:
\begin{align}
\frac{ds}{dt} &= -\frac{\beta_x}{N} sx -\frac{\beta_y}{N} sy,\\
\frac{dx}{dt} &= \frac{\beta_x}{N} sx + \frac{\eta}{N}xy,\\
\frac{dy}{dt} &= \frac{\beta_y}{N} sy - \frac{\eta}{N}xy,\\
N &= s + x + y.
\end{align}

\begin{figure}[!t]
\centering \includegraphics[width=7.56cm]{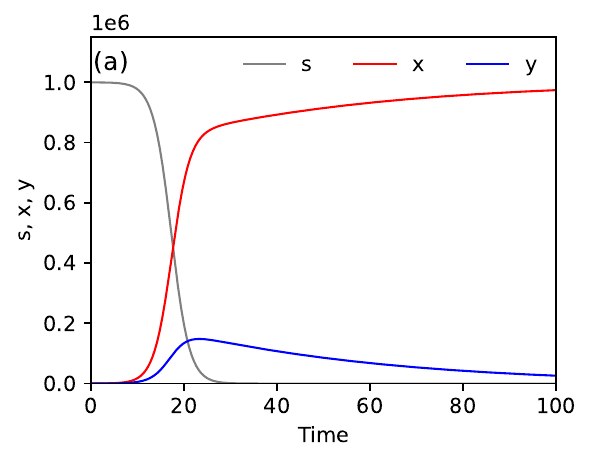}
\centering \includegraphics[width=7.56cm]{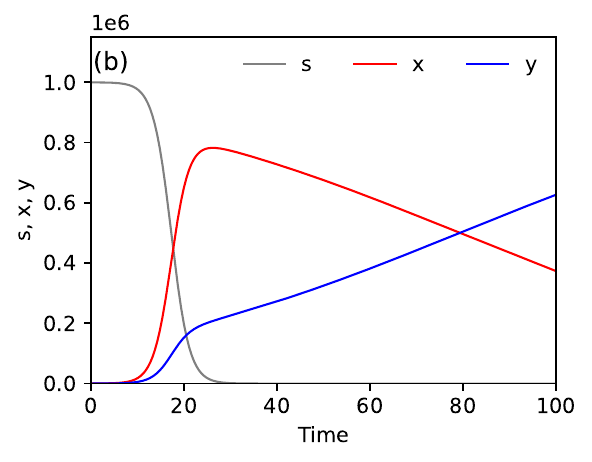}
\caption{Simulation results for the deterministic betrayal model with no regret. The parameters are $T=100$, $N=10^6$, $x_0=100$, $y_0=50$, $\beta_x=0.525$, $\beta_y=0.475$, $\delta_x=0$, $\delta_y=0$ and: (a) $\eta=0.025$; (b) $\eta=-0.025$.}
\end{figure}

Excluding the trivial steady state solution $(x_*,y_*,s_*)=(0,0,N)$, the system has other two possible steady states:
\begin{align}
x_*^{(1)} = N, \quad y_*^{(1)} = 0, \quad s_*^{(1)} = 0,
\end{align}
and respectively
\begin{align}
x_*^{(2)} = 0, \quad y_*^{(2)} = N, \quad s_*^{(2)} = 0.
\end{align}

As previously noticed, none of these steady states can be actually reached, unless the candidates vote against themselves. 
Also, we observe again that these steady states do not depend on the initial condition, 
but as before the speed of approaching the steady state does depend on the initial condition, and therefore it may influence the election outcome in the fixed time $T$ of the campaign.  
In addition, we should note that since the regret rates are equal to zero, the susceptible population also falls to zero, and as a consequence everybody votes. 

In Figure 4 we give a couple of examples to illustrate this case. Here the parameters are $N=10^6$, $x_0=100$, $y_0=50$, $\beta_x=0.525$, $\beta_x=0.475$, $\delta_x=\delta_y=0$. 
If the $y$ candidate is betrayed with the parameter $\eta=0.025$, then the $x$ candidate always wins by a large gap, as shown in Figure 4(a). 
However, if the $x$ candidate is betrayed with $\eta=-0.025$, then the $x$ candidate will unexpectedly lose in the end of the campaign, 
even though it was in the lead most of the time, as shown in Figure 4(b). 
Thus, as expected the betrayal can drastically change the results towards the end of the campaign, even though the candidate had excellent initial chances to win.

\begin{figure}[!t]
\centering \includegraphics[width=7.56cm]{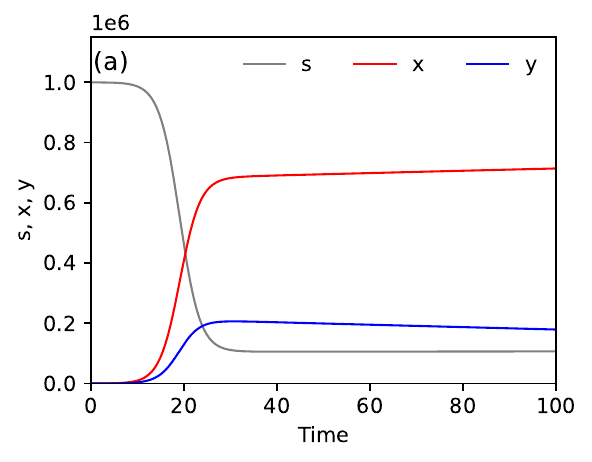}
\centering \includegraphics[width=7.56cm]{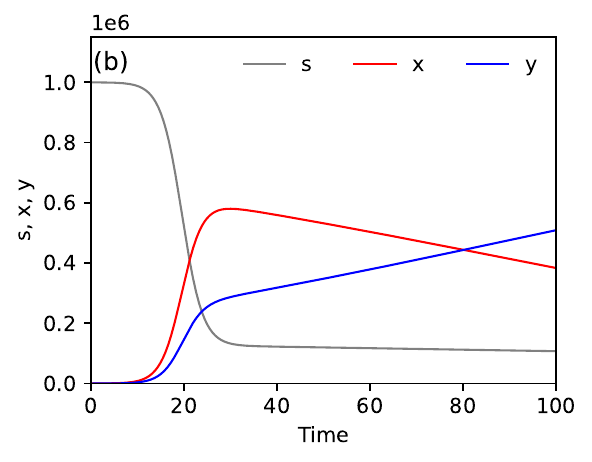}
\caption{Simulation results for the deterministic model with betrayal and regret.
The parameters are $T=100$, $N=10^6$, $x_0=100$, $y_0=50$, $\beta_x=0.525$, $\beta_y=0.475$, $\eta=0.025$, $\delta_y=0.035$ and: (a) $\delta_x=0.06$; (b) $\delta_x=0.077$.}
\end{figure}

\subsection{Betrayal and regret}

This is the general case corresponding to the equations (4-7), when both betrayal and regret are present in the system: $\eta \neq 0$, $\delta_x>0$, $\delta_y>0$. 
In this case, besides the trivial steady state $(x_*,y_*,s_*)=(0,0,N)$, the system also has the following two steady states:
\begin{align}
x_*^{(1)} = \left( 1- \frac{\delta_x}{\beta_x} \right) N, \quad y_*^{(1)} = 0, \quad s_*^{(1)} = \frac{\delta_x}{\beta_x} N,
\end{align}
and respectively
\begin{align}
x_*^{(2)} = 0, \quad y_*^{(2)} = \left( 1- \frac{\delta_y}{\beta_y} \right) N, \quad s_*^{(2)} = \frac{\delta_y}{\beta_y} N,
\end{align}
which also cannot be actually reached, unless the candidates vote against themselves. Also, these states do not depend on the values of initial condition. 
But as previously explained, the speed of approaching the steady states is affected by the initial condition. 

In Figure 5 we give a couple of examples with the parameters: $N=10^6$, $x_0=100$, $y_0=50$, $\beta_x=0.525$, $\beta_x=0.475$, $\delta_y=0.035$. 
In Figure 5(a) the regret parameter is $\delta_x=0.06$, however $x$ is winning with a comfortable gap, due to the betrayal of $y$, even though the population regrets this choice, 
since $\delta_x>\delta_y$. 
In Figure 5(b) the $x$ regret is increased to $\delta_x=0.077$, and the $x$ candidate is losing the election, even though the $y$ candidate is still betrayed with the same rate. 

Thus, without actually knowing the values of the parameters it is quite difficult to provide a coherent explanation of what really happened during the election campaign. 
This observation suggests that several campaign polls/surveys could be useful in estimating the values of the parameters, and to adjust the campaign strategy accordingly.

\section{Stochastic model}

\subsection{Gillespie SSA}

Our approach is based on the Gillespie stochastic simulation algorithm (SSA), initially developed for simulating chemical reactions \cite{key-12}. 
The SSA simulates the time evolution of a chemical reactions system by determining the probabilities of the reactions to occur, and the resulting changes in the 
number of molecules according to the stoichiometry of the reactions. 

Consider a system composed of $J$ chemical species $X_{j}$ $(j=1,...,J)$,
interacting through $M$ reactions $R_{\mu}$ $(\mu=1,...,M)$, 
characterized by their rate constant $k_{\mu}$. The product
$k_{\mu}dt$ is the probability that the reaction $R_{\mu}$
happens in the next infinitesimal time interval $dt$. 

The main steps of the Gillespie algorithm consist of: 
\begin{enumerate}
\item calculating the waiting time $\tau$ for the next reaction to occur; 
\item determining which reaction $\mu$ will occur. 
\end{enumerate}
These quantities are computed by generating two random numbers, $\tau$ and $\mu$, according
to the following probability density function:
\begin{equation}
P(\tau,\mu)=a_{\mu}\exp\left( -\tau\sum_{\mu=1}^{M}a_{\mu}\right),
\end{equation}
where 
\begin{equation}
a_{\mu}=k_{\mu}m_{\mu},
\end{equation}
is the propensity of the reaction $R_\mu$. Here, $m_{\mu}$ is the number of distinct reactant combinations available
for the reaction $R_{\mu}$ in the given state of the system. After the parameters $\tau$ and $\mu$ are generated, 
the numbers of molecules in the system are adjusted according to the stoichiometry of the reaction $R_{\mu}$, and  
the time $t$ is advanced to $t+\tau$.
Below we give the main steps of the algorithm:
\begin{enumerate}

\item Set initial numbers of molecules, set time $t\leftarrow0$.

\item Calculate the propensities, $a_{\mu}$, for all $\mu=1,...,M$.

\item Select $\mu$ with the probability:
\begin{equation}
\Pr(\mu)=\frac{a_{\mu}}{\sum_{\mu=1}^{M}a_{\mu}}.
\end{equation}

\item Select $\tau$ with the probability:
\begin{equation}
\Pr(\tau)=\left(\sum_{\mu=1}^{M}a_{\mu}\right)\exp\left(-\tau\sum_{\mu=1}^{M}a_{\mu}\right).
\end{equation}

\item Change the number of molecules according to the stoichiometry of the reaction $\mu$.

\item Set $t\leftarrow t+\tau$, and go to step 2.

\end{enumerate}

\subsection{The SSA model}

In our case, the implementation of the SSA is simplified by the fact that 
we only have three species $X,Y,S$ corresponding to the populations $x,y,s$. 
Also, initially the number of individuals in each species are: 
\begin{equation}
[X]=X_0, [Y]=Y_0, [S]=N-X_0-Y_0. 
\end{equation}

The "persuasion reactions" can be written as following:
\begin{align}
S + X &\xrightarrow{a_1} 2X, \\
S + Y &\xrightarrow{a_2} 2Y,
\end{align}
with the propensities:
\begin{align}
a_1 &= \frac{\beta_x}{N}[S][X],\\
a_2 &= \frac{\beta_y}{N} [S][Y],
\end{align}
where $[X]$, $[Y]$ and $[S]$ are the number of individuals from the corresponding species at the time $t$. 

The population updates according to the stoichiometry of these reactions are given by:
\begin{align}
[S] &\rightarrow [S] - 1, \quad [X] \rightarrow [X] + 1,\\
[S] &\rightarrow [S] - 1, \quad [Y] \rightarrow [Y] + 1.
\end{align}

The "betrayal reactions" can be written as following:
\begin{align}
X + Y &\xrightarrow{a_3} 2X, \\
X + Y &\xrightarrow{a_4} 2Y,
\end{align}
with the propensities:
\begin{align}
a_3 &= \frac{\eta_x}{N}[X][Y],\\
a_4 &= \frac{\eta_y}{N} [X][Y],
\end{align}
where $\eta_x$ and $\eta_y$ are the betrayal rates, such that the net betrayal rate is $\eta=\eta_y-\eta_x$.

In this case. the $X$ and $Y$ population changes according to the stoichiometry of the reactions as following:
\begin{align}
[Y] &\rightarrow [Y] - 1, \quad [X] \rightarrow [X] + 1,\\
[X] &\rightarrow [X] - 1, \quad [Y] \rightarrow [Y] + 1.
\end{align}

Finally, the "regret reactions" are given by:
\begin{align}
X &\xrightarrow{a_5} S, \\
Y &\xrightarrow{a_6} S,
\end{align}
with the propensities:
\begin{align}
a_5 &= \delta_x [X],\\
a_6 &= \delta_y [Y],
\end{align}
where $\delta_x$ and $\delta_y$ are the regret rates of the voters.
Also, the population changes imposed by the stoichiometry are:
\begin{align}
[X] &\rightarrow [X] - 1, \quad [S] \rightarrow [S] + 1,\\
[Y] &\rightarrow [Y] - 1, \quad [S] \rightarrow [S] + 1.
\end{align}

\subsection{Numerical results}

In a first example (Figure 6), we use SSA (continuous lines) to reproduce the results of the deterministic simulation (dotted lines) from Figure 5. 
Here we used the betrayal rates $\eta_x=0.015$ and $\eta_x=0.04$, such that the net betrayal rate is $\eta=\eta_y-\eta_x=0.025$.
One can see that the SSA simulation follows the deterministic results quite faithfully. 

In general, the agreement between the deterministic and the SSA simulations is really good if the population is large, and the size of the initial "core" support is 
also substantial. However, if the population or the initial "core" support are relatively small, then the stochastic effects induced by the finite size of the population can lead to unexpected results. 

In order to illustrate the above observation, let us consider a relatively smaller community with $N=10^3$ individuals, 
where there are two candidates competing in an election. Initially both candidates lack a "core" support, which means that the initial condition is: $X_0=Y_0=1$, $S_0=998$.
The rest of the parameters are the same as in Figure 6(a): $T=100$, $\alpha=0.525$, $\beta=0.475$, $\eta_x=0.015$, $\eta_y=0.04$, $\delta_x=0.06$, $\delta_y=0.035$. 

The results for two separate runs of the campaign are shown in Figure 7. 
The deterministic model (dotted lines) predicts a comfortable win for the $X$ candidate, 
which does happen in the first run Figure 7(a). However, in a second run Figure 7(b), the $X$ candidate is losing unexpectedly due to the stochastic effects amplified by the fluctuations induced in a smaller population, 
even though the model parameters are the same as in the first run. 

\begin{figure}[!t]
\centering \includegraphics[width=7.56cm]{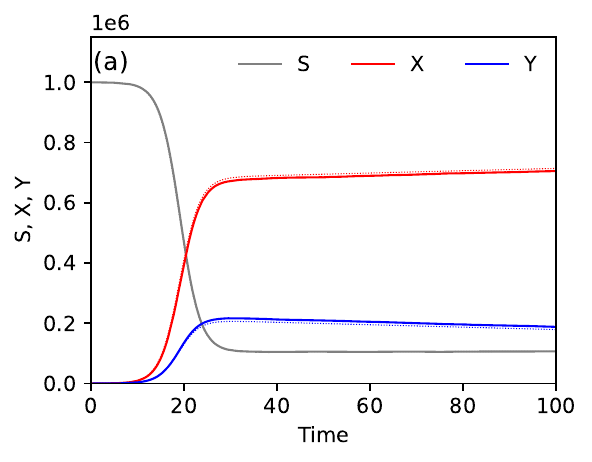}
\centering \includegraphics[width=7.56cm]{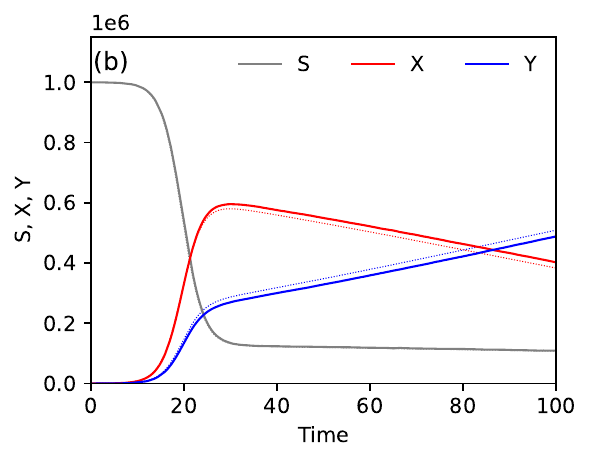}
\caption{SSA simulation with the same parameters as in Figure 5: continuous lines (SSA), dotted lines (deterministic).}
\end{figure}

\begin{figure}[!t]
\centering \includegraphics[width=7.56cm]{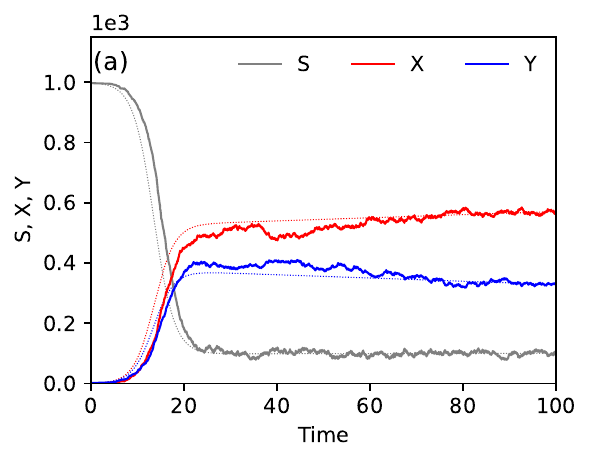}
\centering \includegraphics[width=7.56cm]{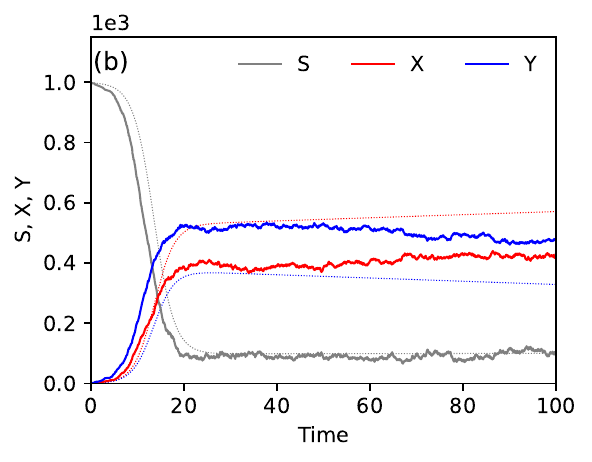}
\caption{Two contradictory results of the SSA simulation with the parameters: $N=10^3$, $X_0=Y_0=1$, $S_0=998$, $\beta_x=0.525$, $\beta_y=0.475$, $\eta_x=0.015$, $\eta_y=0.04$, $\delta_x=0.06$, $\delta_y=0.035$ 
(continuous lines (SSA), dotted lines (deterministic)}. 
\end{figure}

An even more dramatic result is shown in Figure 8. Here we have a large population of $N=10^6$ with the parameters 
$T=100$, $X_0=100$, $Y_0=75$, $\alpha=0.525$, $\beta=0.475$, $\eta_x=0.015$, $\eta_y=0.04$, $\delta_x=0.21$, $\delta_y=0.17$. 
The deterministic model (the dotted lines) predicts that the $X$ candidate will win with a "comfortable" gap of around $4\%$. 
In the first SSA run it does win but in a very tight race, and with a gap of only about $1\%$ (Figure 8(a)). 
In a second SSA run, with the same parameter values, the $X$ candidate is actually losing with a gap of about $3\%$ (Figure 8(b)).  
This happens because of the larger fluctuations induced by the increased voting absence, which is close to $40\%$. 
Therefore, since usually an election happens in a single trial, the results can be strongly influenced by the stochastic nature of the process, if the number of non voting 
individuals increases significantly. 

\begin{figure}[!t]
\centering \includegraphics[width=7.56cm]{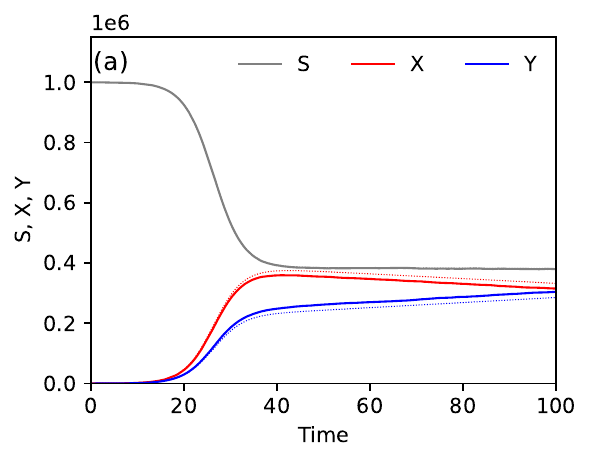}
\centering \includegraphics[width=7.56cm]{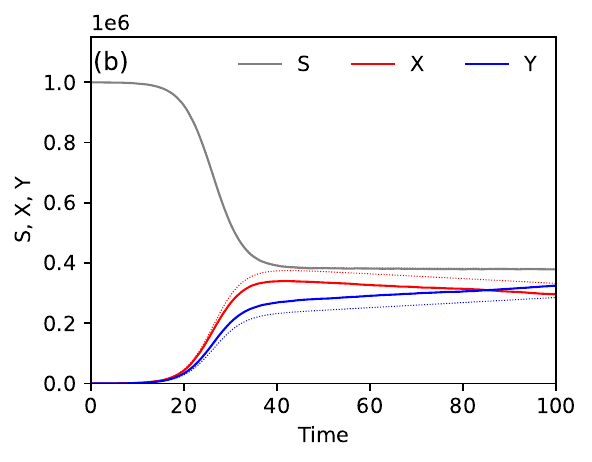}
\caption{SSA simulation illustrating unexpected results due to stochastic effects in a tight race: continuous lines (SSA), dotted lines (deterministic). }
\end{figure}

The number of such dramatic contradictory results decreases fast in the case of larger communities, or for a larger initial "core" support, as shown by the following experiment (Figure 9) where we use the same parameters as before 
($T=100$, $\alpha=0.525$, $\beta=0.475$, $\eta_x=0.015$, $\eta_y=0.04$, $\delta_x=0.06$, $\delta_y=0.035$), but we increase the initial "core" support for the candidates as following: 
(a) $X_0=Y_0=1$, (b) $X_0=Y_0=5$, (c) $X_0=Y_0=10$, (d) $X_0=Y_0=15$. The results are averaged over 1000 simulated elections. 
One can see that the percentage of wins for the $X$ candidate increases from about $55\%$ to about $95\%$. 

\begin{figure}[!h]
\centering \includegraphics[width=7.56cm]{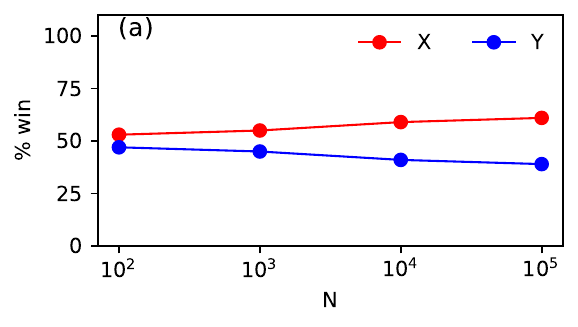}
\centering \includegraphics[width=7.56cm]{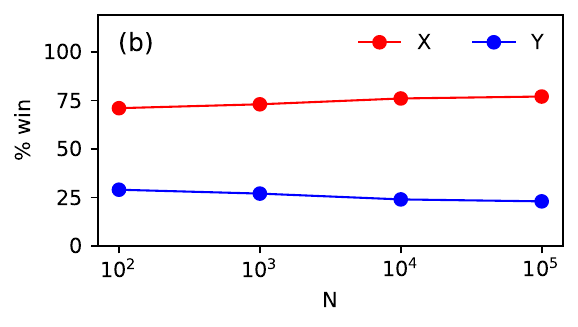}
\centering \includegraphics[width=7.56cm]{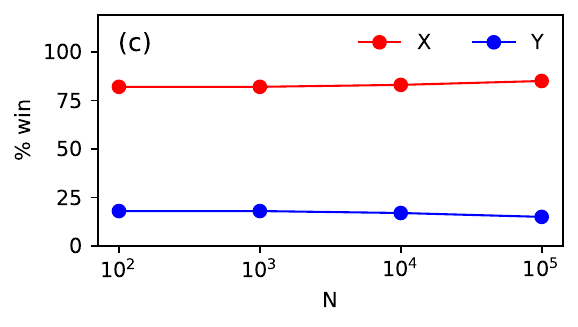}
\centering \includegraphics[width=7.56cm]{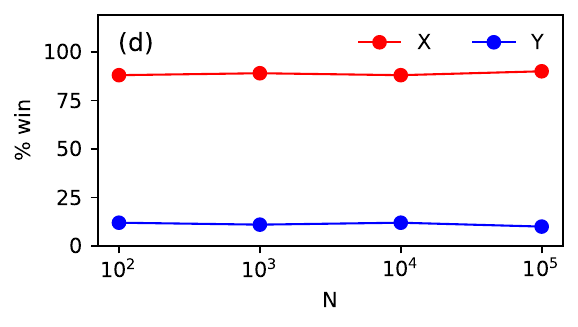}
\caption{Percentage of wins if the initial support is the same for both candidates.} 
\end{figure}

\begin{figure}[!h]
\centering \includegraphics[width=7.56cm]{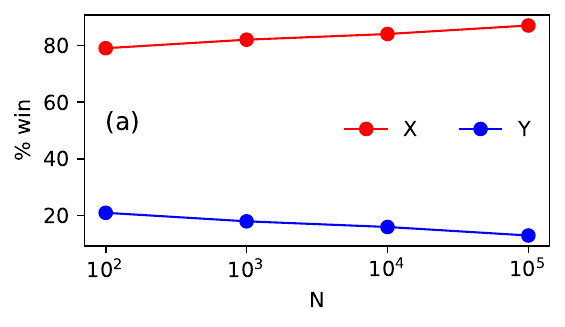}
\centering \includegraphics[width=7.56cm]{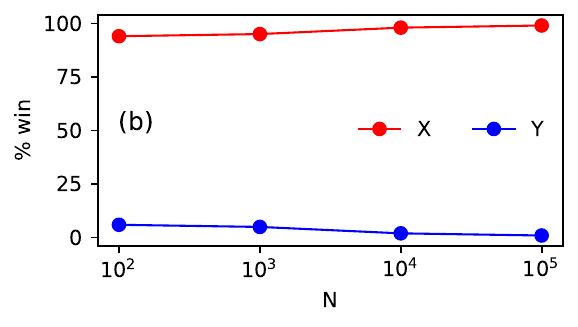}
\centering \includegraphics[width=7.56cm]{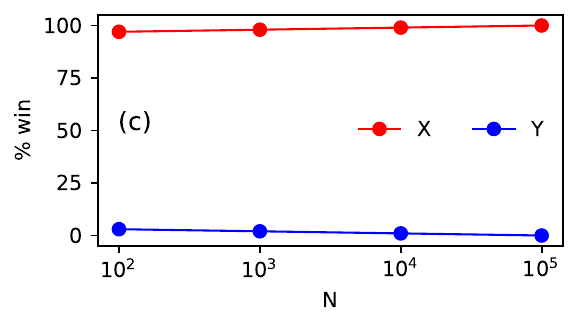}
\centering \includegraphics[width=7.56cm]{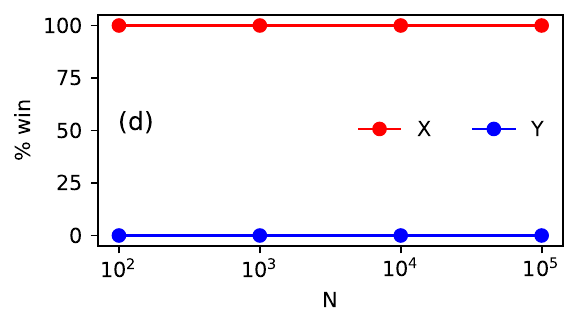}
\caption{Percentage of wins if the initial "core" support is doubled for candidate $X$.}
\end{figure}

In the previous numerical experiment, the initial "core" support had the same value for both candidates, which makes the results more susceptible to fluctuations at lower population levels, or for lower "core" support values. 
However, if the initial "core" support is maintained the same for the $Y$ candidate, but it is  doubled for the $X$ candidate, then the $X$ candidate will win with a high probability as shown in Figure 10 for: 
(a) $X_0=2, Y_0=1$, (b) $X_0=10, Y_0=5$, (c) $X_0=20, Y_0=10$, (d) $X_0=30, Y_0=15$.

\section*{Conclusion}

We have discussed a novel election campaign model inspired by the compartmental epidemiology, and 
we have shown that the model captures some of the the main characteristics of an election campaign: persuasion, betrayal and regret. 
Also, we have presented detailed cases exemplifying how these three characteristics of the model can be used to influence the outcome of the campaign, using both deterministic and stochastic simulations. 
Moreover, the simulations show that the decision to not vote (abstention) significantly increases the fluctuations in the model, amplifying the chance of controversial results.  
The deterministic models are unable to capture such "finite size effects" induced by the fluctuations in a smaller population size, or by a smaller initial "core" support, 
or due to very tight racing, and in such cases stochastic simulations should be preferred. 

An extension of the deterministic and stochastic models to a larger number of candidates is provided in the Appendix A. Also, the Python code used to perform the simulations 
from the paper is provided in the Appendix B.

\section*{Appendix A}

\subsection*{Extended deterministic model}

Here we generalize the deterministic model for $M\geq 2$ candidates. We assume a population of $N$ individuals, where $M$ of them are candidates for an election. 
We denote by $x_m$ the population supporting the candidate $m \in \{1,2,...,M\}$, and by $s$ the susceptible population. The initial condition at $t=0$ is denoted by 
$\{x_1(0),x_2(0),...,x_M(0)\}$ and it represents the initial "core" support for the candidates, such that the initial susceptible population is:
\begin{equation}
0<s(0) = N-\sum_m x_m(0) < N.
\end{equation}
The duration of the election campaign is fixed 
to a time interval $[0,T]$. We also denote by $\beta_m$ the persuasion rates, and by $\delta_m$ the regret rates. Also, $\eta_{mn}$ is the net betrayal rate of the candidate 
$n$ in favor of candidate $m$, such that $\eta_{nm} = -\eta_{mn}$ for $m\neq n$, and $\eta_{mm}=0$. We should note that the matrix $\eta$ should be very sparse, and typically only one element 
per row is non zero, since in general one candidate is betrayed in favor of another candidate only. 
With this notation the generalized deterministic model consists of the following $M+2$ differential equations:
\begin{align}
\frac{ds}{dt} &= \sum_{m} \left(  -\frac{\beta_m}{N} sx_m  + \delta_m x_m \right) ,\\
\frac{dx_m}{dt} &= \frac{\beta_m}{N} sx_m - \delta_m x_m + \sum_{n\neq m} \frac{\eta_{mn}}{N}x_m y_n\\
N &= s + \sum_{m} x_m.
\end{align}
The last equation is the total population conservation constraint, and it normalizes the results to the total number of voters.

\subsection*{Extended stochastic model}

We use the same notation as above in order to generalize the stochastic model to $M\geq 2$ candidates. Also, 
the initial number of individuals in each species are $[X_m(0)]$, $m=1,2,...,M$, such that the initial susceptible population is given by:
\begin{equation}
0<[S(0)] = N-\sum_m [X_m(0)] < N. 
\end{equation}

The "persuasion reactions" can be written as following:
\begin{align}
S + X_m &\xrightarrow{a_m} 2X_m, 
\end{align}
with the propensities:
\begin{align}
a_m &= \frac{\beta_m}{N}[S][X_m].
\end{align}
The population updates according to the stoichiometry of these reactions are given by:
\begin{align}
[S] &\rightarrow [S] - 1, \quad [X_m] \rightarrow [X_m] + 1.
\end{align}

The "betrayal reactions" can be written as following:
\begin{align}
X_m + X_n &\xrightarrow{a_{mn}} 2X_m,
\end{align}
with the propensities:
\begin{align}
a_{m} &= \frac{\eta'_{mn}}{N}[X_m][X_n],
\end{align}
where $\eta'_{mn}$ is the absolute betrayal rate (the net betrayal rate is: $\eta_{mn}=\eta'_{nm}-\eta'_{mn}$). 
The population changes according to the stoichiometry of the reactions as following:
\begin{align}
[X_n] &\rightarrow [X_n] - 1, \quad [X_m] \rightarrow [X_m] + 1.
\end{align}

The "regret reactions" are given by:
\begin{align}
X_m &\xrightarrow{a'_m} S, 
\end{align}
with the propensities:
\begin{align}
a'_m &= \delta_m [X_m]. 
\end{align}
Also, the changes imposed by the stoichiometry are:
\begin{align}
[X_m] &\rightarrow [X_m] - 1, \quad [S] \rightarrow [S] + 1.
\end{align}

\section*{Appendix B}

\subsection*{Numerical implementation of the deterministic model}

In order to implement numerically the deterministic version of the model we use Python with Numpy and Scipy packages, and Matplotlib for visualization. 
From Scipy we use the "odeint" function from the "integrate" class, which can be used to solve a system of ordinary differential equations. 
This "integrate" class implements an implicit multistep method using the backward differentiation formula, specially developed for handling 
both stiff and nonstiff systems of differential equations \cite{key-13}. The Python code of the deterministic model implementation is given below:

\begin{verbatim}
import numpy as np
from scipy import integrate
from matplotlib import pyplot as plt

def derivative(X, t):
    s, x, y = X
    ds = -beta_x*s*x/N - beta_y*s*y/N + delta_x*x + delta_y*y
    dx = beta_x*s*x/N - delta_x*x + eta*x*y/N
    dy = beta_y*s*y/N - delta_y*y - eta*x*y/N
    return np.array([ds, dx, dy])
    
def plot_results(fig_name, t, s, x, y):
    fig = plt.figure(figsize=(4,3))
    plt.xlabel("Time") 
    plt.ylabel("s, x, y")
    plt.xlim([0, T]) 
    plt.ylim([0, 1.15*N])
    plt.plot(t, s, c="grey", linewidth=1.0, label="s")
    plt.plot(t, x, c="red", linewidth=1.0, label="x")
    plt.plot(t, y, c="blue", linewidth=1.0, label="y")
    plt.legend(ncol=3, frameon=False)
    plt.gca().ticklabel_format(axis='y', style='sci', scilimits=(0, 0)) 
    fig.savefig(fig_name, bbox_inches='tight')

if __name__ == "__main__":
    N, x0, y0, T = 1e6, 100, 50, 100
    s0 = N - x0 - y0
    beta_x, beta_y, eta = 0.525, 0.475, 0.025
    delta_x,delta_y = 0.077,0.035
    t = np.linspace(0, T, T+1) 
    
    X = s0, x0, y0 #initial conditions vector
    s, x, y = integrate.odeint(derivative, X, t).T
    plot_results("fig.pdf", t, s, x, y)
\end{verbatim}

\subsection*{Numerical implementation of the stochastic model}

The stochastic version of the model was implemented from scratch using Python with the Numpy package, and Matplotlib for visualization. 
The "ssa" function implements the Gillespie SSA method. The input to the function is a dictionary ("model") containing the 
initial count values of the population, the duration of the simulation, the stoichiometry of the reactions (as an array where each row corresponds to a reaction), 
and the propensities of the reactions (computed using an array of lambda functions). The rows in the propensities array correspond to the 
rows in the stoichiometry array. 
We should emphasize that since this is a stochastic simulation the results will be different for each run. 
The code of the Python implementation is given below:

\begin{verbatim}
import numpy as np
from matplotlib import pyplot as plt

def ssa(model):
    t, c = [0.0], [model["counts"]]
    while t[-1] < model["duration"]:
        a = np.array([q(*c[-1]) for q in model["propensities"]])
        w = np.sum(a)
        if w == 0:
            break
        r = np.random.random(2)       
        j, h = 0, r[0]*w - a[0]
        while h > 0:
            j += 1
            h -= a[j]
        q = model["stoichiometry"][j]
        s = [x + y for x,y in zip(c[-1],q)]
        t.append(t[-1] - np.log(r[1])/w)
        c.append(s)       
    t.append(model["duration"])
    c.append(c[-1])
    return np.array(t), np.array(c)

def plot_results(fig_name, t, c):
    fig = plt.figure(figsize=(4,3))
    plt.xlabel("Time")
    plt.ylabel("s, x, y")
    plt.xlim([0, T])
    plt.ylim([0, 1.15*N])
    plt.plot(t, c[:,0], c="grey", linewidth=1.0, label="s")
    plt.plot(t, c[:,1], c="red", linewidth=1.0, label="x")
    plt.plot(t, c[:,2], c="blue", linewidth=1.0, label="y")
    plt.legend(ncol=3, frameon=False)
    plt.gca().ticklabel_format(axis='y', style='sci', scilimits=(0, 0)) 
    fig.savefig(fig_name, bbox_inches='tight')

if __name__ == '__main__':
    N, x0, y0, T = 1e6, 100, 50, 100
    s0 = N - x0 - y0
    beta_x, beta_y = 0.525, 0.475
    eta_x, eta_y = 0.015, 0.04
    delta_x, delta_y = 0.06, 0.035
    model = {
        "counts": [s0, x0, y0],
        "duration": T, 
        "stoichiometry": [[-1, 1, 0],
                          [-1, 0, 1],
                          [0, -1, 1],
                          [0, 1, -1],
                          [1, -1, 0],
                          [1, 0, -1]],
        "propensities": [lambda s, x, y: beta_x * s * x/N,
                         lambda s, x, y: beta_y * s * y/N,
                         lambda s, x, y: eta_x * x * y/N,
                         lambda s, x, y: eta_y * x * y/N,
                         lambda s, x, y: delta_x * x,
                         lambda s, x, y: delta_y * y]
    }        
    t, c = ssa(model)
    plot_results("fig.pdf", t, c)  
\end{verbatim}

\end{document}